\documentclass[aps, amsmath,twocolumn,superscriptaddress,floatfix]{revtex4-2}

\usepackage{graphicx}% Include figure files
\usepackage{dcolumn}% Align table columns on decimal point
\usepackage{bm}% bold math
\usepackage{color}

\newcommand{\modif}[1]{\textcolor{black}{#1}}

\begin{document}

\preprint{APS/123-QED}

\title{Magnetic domain depinning as possible evidence for two ferromagnetic phases in LaCrGe$_3$}

\author{R. R. Ullah}
\address{Department of Physics and Astronomy, University of California, Davis, California 95616, USA}
\author{P. Klavins}
\address{Department of Physics and Astronomy, University of California, Davis, California 95616, USA}
\author{X. D. Zhu}
\address{Department of Physics and Astronomy, University of California, Davis, California 95616, USA}
\address{Department of Optical Sciences and Engineering, Fudan University, Shanghai 200045, China}
\author{V. Taufour}
\address{Department of Physics and Astronomy, University of California, Davis, California 95616, USA}

\date{April 29, 2023}% It is always \today, today,
             %  but any date may be explicitly specified

\begin{abstract}
Two ferromagnetic phases, FM1 and FM2, were first proposed to exist in LaCrGe$_3$ based on a broad maximum in the temperature derivative of resistivity resembling that of the superconducting ferromagnet UGe$_2$ where FM1 and FM2 are well-established. While evidence for two FM phases can be found in certain additional probes, corresponding anomalies in magnetization have not been recognized until now. Our spatially-resolved images of the magnetic domains show a substantial change in the domain structure between the higher temperature FM1 phase and the lower temperature FM2 phase. Furthermore, our measurements of the coercive field and virgin magnetization curves reveal an unconventional magnetic domain pinning region in the FM1 phase, followed by a depinning region at lower temperatures where the system is reported to crossover into the FM2 phase. We incorporate this discovery into a simple domain magnetization model that demystifies the magnetization curve seen in all previous studies. Finally, we find that the unusual domain behavior can be explained by a change in the ferromagnetic exchange interaction and magnetic moment, both of which are consistent with the existence of two FM phases. This revelation may help explain a range of anomalous behaviors observed in LaCrGe$_3$ and rekindles the discussion about the prevalence of multiple FM phases in fragile FM systems.
\end{abstract}

%

%\keywords{Suggested keywords}%Use showkeys class option if keyword
                              %display desired
\maketitle

%\tableofcontents
\section{\label{sec:Intro}Introduction\protect\\}
The ferromagnetic compound LaCrGe$_3$ ($T_\mathrm{C} = 85$\,K, BaNiO$_3$-type crystal structure) is a proven playground for studying the suppression of ferromagnetism under pressure. Many multiprobe experiments have mapped its temperature-pressure-magnetic field phase diagram~\cite{Taufour2016, Kaluarachchi2017, Taufour2018, Rana2021, Gati2021}. These experiments exploring ferromagnetic quantum criticality in LaCrGe$_3$ take place in extreme conditions and, therefore, leave unsolved mysteries. For example, the magnetic phase that appears in place of a quantum critical point was initially proposed to be an AFM$_Q$ phase~\cite{Taufour2016, Taufour2018} in agreement with the theoretical proposals of avoided ferromagnetic quantum critical points~\cite{Belitz1997PRB,Chubukov2004PRL,Conduit2009PRL,Karahasanovic2012PRB,Thomson2013PRB,Pedder2013PRB,Belitz2017PRL,Wysokinski2019SR}, but the Q-vector has proven difficult to determine, and recently, a short-ranged FM order phase has been suggested as an alternative~\cite{Gati2021}. The mystery relevant to our study, however, is the one surrounding the existence of two ferromagnetic states, FM1 and FM2, within the FM region.

\begin{figure}[!htb]
\includegraphics[width=8.6cm]{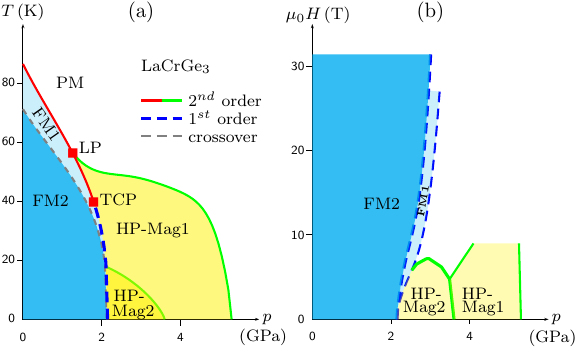}
\caption{\label{fig:FM1FM2ab} (a) Temperature-pressure and (b) magnetic field-pressure phase diagrams of LaCrGe$_3$ deduced from resistivity, magnetization, $\mu$SR measurements~\cite{Taufour2016,Kaluarachchi2017,Taufour2018}. The paramagnetic (PM), high-pressure magnetic phases (HP-Mag1 and HP-Mag2) and the two ferromagnetic phases (FM1 and FM2) are indicated. The Lifshitz point (LP) and tricritical point (TCP) are also shown. (a) At zero field, a broad maximum in the temperature derivative of resistivity has been interpreted as a crossover between two ferromagnetic phases FM1 and FM2. Under pressure, the crossover line merges with the first-order transition line between FM and HP-Mag2. (b) Under field at low temperature there are successive field-induced first-order transitions into the FM1 and FM2 states.}
\end{figure}

The potential for two ferromagnetic phases in LaCrGe$_3$ was first realized based on a broad maximum in the temperature derivative of resistivity, $d\rho_{ab}/dT$~\cite{Kaluarachchi2017}, that resembled that in UGe$_2$, where two ferromagnetic phases are well established~\cite{Tateiwa2001JPSJ, Pfleiderer2002, Hardy2009PRB, Taufour2010}. In UGe$_2$, the phases have distinct moments and the magnetic ground state evolves from FM2 (M$_0 \approx$ 1.5\,$\mu_B/$U) to FM1 (M$_0\approx$ 0.9\,$\mu_B/$U) at pressure $p_x \approx 1.2$\,GPa, before becoming paramagnetic above $p_c\approx 1.5$\,GPa. In LaCrGe$_3$, however, applying pressure causes the crossover boundary between FM1 and FM2 to merge with the quantum phase transition line (i.e. $p_x=p_c$) and FM1 is not accessible at zero temperature, as seen in Fig.~\ref{fig:FM1FM2ab}a. Still, applying magnetic field separates the phases again and features of a phase transition are readily observed and form two planes of first order transitions, called wings, ending at quantum wing critical points~\cite{Kaluarachchi2017}. At ambient pressure, however, there is no genuine phase transition between FM1 and FM2. Instead, there is a crossover regime which is allowed if FM1 and FM2 have the same symmetry, as is the case in UGe$_2$ where the only difference between the two phases is the size of the magnetic moment. In ambient pressure LaCrGe$_3$, evidence for a crossover \modif{at $T_x \approx 70$\,K} has been observed in many physical properties, such as $d\rho_{ab}/dT$~\cite{Kaluarachchi2017}, $d\rho_c/dT$~\cite{Gati2021}, specific heat~\cite{Lin2013}, and thermoelectric power~\cite{Das2014}, but features of FM1 and FM2 have yet to be recognized in magnetization, $M$.

In this article, we present magnetization data on LaCrGe$_3$ that supports the existence of FM1 and FM2. Our spatially-resolved images of the magnetic domains reveal a significant change in the domain structure on either side of the FM1-FM2 crossover. We find that the temperature dependence of the coercive field, $H_c$, shows the rare case in which $H_c$ increases with temperature and reaches a local maximum at $72.5$\,K, \modif{near $T_x \approx 70$\,K} where the crossover between FM1 and FM2 is reported to occur. Furthermore, we observe the unusual situation where the virgin magnetization curve is limited by domain wall pinning at high temperatures ($T\gtrsim60$\,K), but subsequently shows no domain wall pinning at low temperatures ($T\lesssim60$\,K). We are able to incorporate this change between domain-wall immobility and mobility into a simple model that beautifully recreates the previously unexplained features in the magnetization as a function of temperature curves observed in LaCrGe$_3$ at low applied fields. Finally, we show that this difference in domain wall mobility can possibly be caused by a change in the moment and the ferromagnetic exchange constant, both of which are consistent with a crossover between two ferromagnetic states.

There is evidence that two ferromagnetic states are not uncommon in fragile FM systems. In the case of LaCrGe$_3$, we recognized that the FM1-FM2 crossover can be detected in magnetization by way of spatially resolved images as well as bulk DC magnetization measurements. Other measurements also show anomalies in similar temperature regions that may stem from the crossover. For example, our domain-wall pinning and depinning analysis is directly supported by AC susceptibility measurements which indicate domain pinning in the similar temperature region of increased coercivity~\modif{\cite{Bie2009a, BoschSantos2021a}}. Interestingly, the AC susceptibility data shows two peaks, which also appear in two other compounds with ferromagnetic quantum phase transitions, UCoAl~\cite{Kimura2015} and Sr$_3$Ru$_2$O$_7$~\cite{Wu2011}. A sharp peak below $T_\mathrm{C}$ \modif{and near $T_x$} was also observed in recent ac susceptibility measurements with \modif{the} field applied perpendicular to the $c$ axis~\cite{XuArxiv}. Additional evidence for the FM1-FM2 crossover can be found in thermoelectric power measurements where the local minimum attributed to the crossover in UGe$_2$~\cite{Morales2016} also appears in LaCrGe$_3$~\cite{Das2014} at the same temperature where the crossover is reported to occur at ambient pressure. There are also two peaks in the ESR spectra~\cite{Sichelschmidt2021} which can be found in other compounds with two FM states such as La$_{1-x}$Te$_x$MnO$_3$~\cite{Tan2003}. Due to these similarities in features to other FM compounds with rich magnetic phase diagrams, along with our fresh look at features in magnetization, both of which can be explained by multiple ferromagnetic states, we find it unlikely that the FM phase in LaCrGe$_3$ is a simple one.

\section{\label{sec:Methods}Experimental Methods}

\subsection{Crystal Synthesis}
Single crystals of LaCr$\textrm{Ge}_3$ were synthesized with a self-flux solution growth technique~\cite{Lin2013} and characterized by powder x-ray diffraction. Further synthesis details can be found in Appendix~\ref{subsec:synthesis}.
 
\subsection{Bulk Magnetic Measurements}
We use a Quantum Design Magnetic Property Measurement System (MPMS) to measure the magnetization of the sample with the applied magnetic field oriented parallel to the $c$ axis. Magnetization as a function of temperature ($M$\,v\,$T$)  was measured in various constant applied fields and by the following methods: field-cooled-cooling (FCC), field-cooled-warming (FCW), and zero-field-cooling (ZFC). Isothermal magnetization as a function of applied field ($M$\,v\,$H$) was measured at a selection of temperatures below $T_\mathrm{C}$. Additional details regarding these measurement procedures can be found in Appendix~\ref{subsec:AppendixMagDetails}.

\begin{figure*}
\centering
\includegraphics[trim = 15 20 0 0]{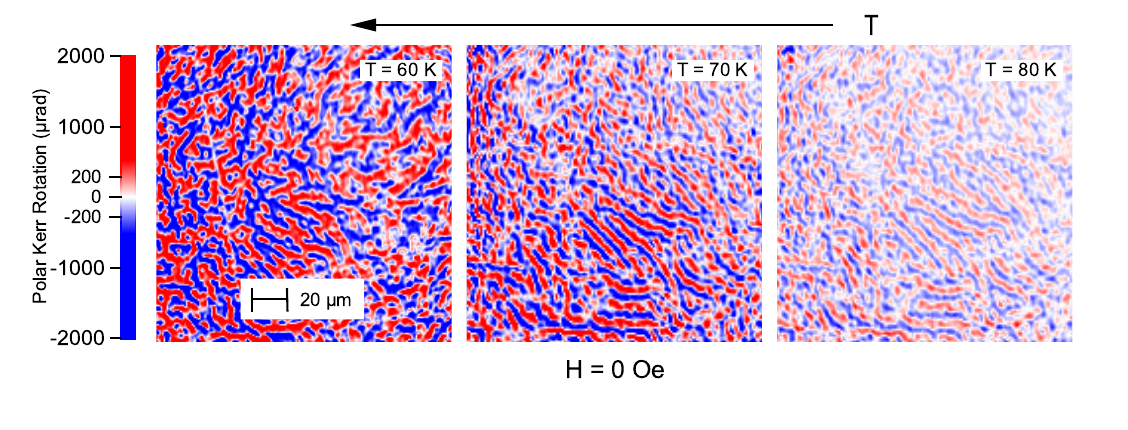}
\caption{\label{fig:MOKE}Polar Kerr rotation images of a polished $ab$ face of LaCrGe$_3$ acquired while cooling in zero applied field at $60$\,K, $70$\,K, and $80$\,K. The scan area shown is $170$$\,\mu$m x $170$$\,\mu$m. Positive values of Kerr rotation are shown in red and indicate a region of magnetization pointing out of the page, while negative values are shown in blue and depict a region of $M$ pointing into the page. The striking similarity of the domain structure between $80$\,K and $70$\,K is evidence for domain wall pinning in that range of temperature. In contrast, we observe a drastic change in the size and shape of the domains when the sample is cooled from $70$\,K to $60$\,K. The depinning of domain walls upon cooling is unusual, and may be due to the crossover between FM1 and FM2.}
\end{figure*}
\subsection{Magnetic Domain Imaging}
Polar \modif{magneto-optical Kerr effect (MOKE)} images of the $ab$ face of a LaCrGe$_3$ single crystal are acquired using a normal-incidence Sagnac interferometric scanning microscope~\cite{Zhu2022RSI, Zhu2017, Zhu2021a}. The microscope has a spatial resolution of 0.85\,$\mu$m and a sensitivity of 0.4\,$\mu$rad. The sample is placed in an optically accessible flow cryostat and its temperature can be varied from $470$\,K to $11.2$\,K. Longitudinal MOKE images of the $ac$ face of a LaCrGe$_3$ single crystal are acquired with an oblique-incidence Sagnac interferometric scanning microscope \cite{Zhu2021RSI} with a spatial resolution of $25$\,$\mu$m.

Since the LaCrGe$_3$ single crystal naturally grows along its $c$ axis, the $ab$ face is obtained by polishing with fine 0.1\,$\mu$m particle size Al$_2$O$_2$ polishing papers. Bulk magnetization was measured before and after polishing and no magnetic impurities potentially picked up from polishing were observed.

Optical images of magnetic domains on an as-grown $ac$ face of a LaCrGe$_3$ single crystal were reported in an earlier study using an oblique-incidence Sagnac interferometric scanning microscope down to $77.4$\,K~\cite{Zhu2021RSI}. This study was limited to liquid nitrogen temperatures and thus, domain structures on both sides of the FM1-FM2 crossover were not investigated. Furthermore, because the easy magnetization axis (i.e., the $c$ axis) lies in the $ac$ plane, an oblique-incidence microscope had to be used in order to image magnetic domains using longitudinal and transverse Kerr rotation effects. In our present study, we image the $ab$ face of the sample using polar Kerr rotation effects, to study magnetic domains with a normal-incidence microscope which has a much better spatial resolution ($0.85$\,$\mu$m compared to $25$\,$\mu$m) and two orders of magnitude better sensitivity compared to an oblique-incidence microscope.

\section{\label{sec:Results}Results and Discussion}
\subsection{Magnetic Domain Structure Imaging}
%For uniaxial ferromagnets, the domains on the ab face will terminate on the surface.
%The larger demagnetization effect will lead to smaller domains in order to minimize the external field. 

In this work, we examine magnetic domains on the $ab$ face down to $20$\,K. The images shown in Fig.~\ref{fig:MOKE} were taken at $80$\,K, $70$\,K, and $60$\,K while cooling down in zero applied field. Between $80$\,K and $70$\,K, the domain structure is unchanged, albeit for an increase in contrast at lower temperature due to the corresponding increase in the spontaneous magnetization, $M_s$. Between $70$\,K and $60$\,K, however, there is a drastic change in domain structure. By comparing the domains at $60$\,K to those at $70$\,K and $80$\,K, we observe that the shape and size of the domains change dramatically.

The lack of change in the domain structure between $80$\,K and $70$\,K supports our hypothesis that the domain walls are pinned in this temperature region which roughly encompasses the FM1 state. The following radical change in the domain structure upon cooling is made possible by the depinning of domain walls between $70$\,K and $60$\,K. Furthermore, the change in size of the domains suggests that the energy cost of the domain wall changed between these two temperatures. This change in domain structure coincides with the crossover between FM1 and FM2 reported to occur in this temperature region.

It is highly unusual for domain wall depinning to occur upon cooling in FM systems, since the thermal energy available to overcome the energy barrier to wall movement is reduced. While the following sections will show how the domain depinning anomaly can be probed with bulk magnetization measurements, these spatially-resolved magneto-optical measurements are particularly suited for observing the change in domain structure. Similar imaging with this MOKE microscope was performed on the magnetic domains in single crystals of the Weyl semimetal Co$_3$Sn$_2$S$_2$~\cite{Shen2022}. Those images supported the idea that the domain walls depin in a narrow range of temperature while cooling down, therefore explaining an anomalous downturn in magnetization feature that had garnered considerable attention~\cite{Kassem2017,Kassem2021,Zhang2021,Soh2022,Hu2022,Shen2019,WU2020,Lachman2020,Zhang2022,Lee2022,Howlader_2021}. A different MOKE study suggested that the underlying cause of the change in domain wall mobility in Co$_3$Sn$_2$S$_2$ is a domain wall transition from linear walls to Bloch due to its unusually large dimensionless anisotropy factor $K$~\cite{Lee2022}. In the case of LaCrGe$_3$, we find $K$ is an order of magnitude smaller than in Co$_3$Sn$_2$S$_2$ and therefore a domain wall transition is unlikely, although spin-polarized low energy electron microscopy (SPLEEM)~\cite{Chen2020} should be performed to confirm. Even without the knowledge of the spin alignment inside the domain walls that SPLEEM would provide, we find the distinct domain structure observed above and below $70$\,K to be convincing evidence for the existence of two ferromagnetic states.

\subsection{\label{sec:Hysteresis}Reappearing Hysteresis Loops and Domain Pinning Virgin Curves}

\begin{figure}%[h]
\includegraphics[trim = 140 10 110 0]{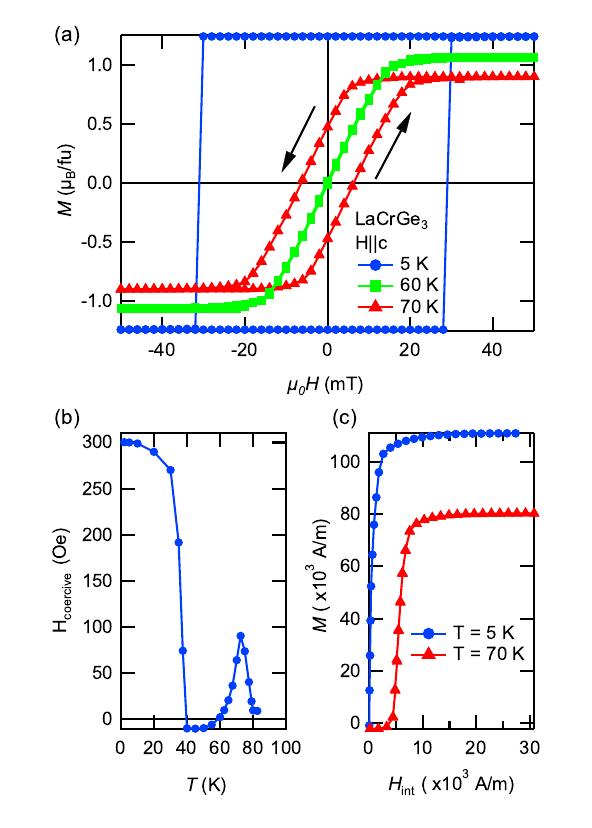}
\caption{\label{fig:Hysteresisfig} (a) Magnetic hysteresis loops measured at different temperatures below the Curie temperature ($T_\mathrm{C}$ = 85\,K). Starting at $2$\,K, the hysteresis loops are rectangular and shrink in size as $T$ increases until they close completely near $40$\,K. Surprisingly, the hysteresis loop opens up again before reaching $T_\mathrm{C}$. (b) The coercive field, $H_c$, as a function of temperature shows the low $T$ and high $T$ regions of coercivity are separated by a region where $H_c\approx0$. The increase of coercive field with temperature in the $67$\,K to $72.5$\,K region is unusual and may be due to a crossover between FM1 and FM2. (c) The virgin magnetization curves reveal an atypical change in domain wall mobility. In the low temperature FM2 state, the virgin magnetization curve follows demagnetization theory which means the domain walls are not pinned. In the higher temperature FM1 state, the hesitance for the virgin curve to increase with field suggests domain wall pinning.}
\end{figure}

From measuring hysteresis loops at different temperatures, we find changes in the loop shape, coercivity, and virgin magnetization curve that are consistent with domain wall pinning in the high temperature FM1 state, followed by depinning in the low temperature FM2 state. At low temperatures, the loops are rectangular, characterized by a remanent magnetization that is nearly equal to $M_s$ and a sudden reversal of the sample magnetization, $M$, at the coercive field, $H_c$. Similar rectangular loops were recently reported~\cite{XuArxiv}. As expected, the hysteresis loops shrink in width with increasing temperature until they fully close. Surprisingly, LaCrGe$_3$ is a rare case in which further increasing the temperature causes the hysteresis loop to reappear, albeit with a much more gradual change in $M$ compared to the low $T$ loops. An example of these three kinds of hysteresis loops can be found in Fig.~\ref{fig:Hysteresisfig}a.

By plotting $H_c$ as a function of temperature, as shown in Fig.~\ref{fig:Hysteresisfig}b, we can see the low $T$ ($T<40$\,K for this particular sample) and high $T$ ($62.5$\,K\,$<T< T_C$ for all samples measured) regions of coercivity, separated by a region where the coercivity is minimal. Our results are consistent with a recent report~\cite{XuArxiv}. In the low $T$ region, the magnitude of $H_c$ as well as the temperature at which $H_c$ disappears, is sample-dependent, which leads us to believe it is related to sample quality and defects \modif{(as discussed further in Appendix~\ref{sec: SampleDependence})}. The coercivity at high $T$, however, is evident in all samples measured, occurring at the same temperatures with similar magnitudes which suggests it is related to an intrinsic property of the compound. This repeatability between samples is consistent with the scenario that the high $T$ coercivity originates from the FM1-FM2 crossover.

A careful analysis of the virgin magnetization curves and their relationship to the shape of the hysteresis loops confirms our domain pinning hypothesis. Examples of the low $T$ and high $T$ curves are shown as the blue dots and red triangles, respectively, in Fig.~\ref{fig:Hysteresisfig}c. At low $T$, the virgin curve increases linearly with a slope of $1/N_c$, where $N_c$ is the demagnetization constant along the $c$ axis (see Appendix~\ref{sec: Demag} for details), and $M$ saturates to $M_s$ at an applied field much lower than $H_c$. This behavior can only occur if the domain walls can move freely. On the other hand, in the high $T$ region of coercivity, the virgin curve barely increases in response to $H_{\textrm{applied}}$ until $H_{\textrm{applied}} = H_c$. This behavior suggests that domain wall pinning is present in this range of temperatures~\cite{Bertotti1998, TremoletdeLacheisserie2004}.

LaCrGe$_3$ is not the first sample to display an $H_c$ that re-emerges with increasing temperature. For example, in Gd, the coercivity is also split between two regions by a minimum $H_c$ below its $T_C = 289$\,K~\cite{Belov1961,Bates1961PPS}. This similarly shaped coercivity plot in Gd, however, is due to a change of sign of the anisotropy constant $K_1$ and comparatively large values of $K_2$~\cite{Corner1962,Graham1963}. In contrast, no anomaly is observed in the temperature dependence of the anisotropy constants in LaCrGe$_3$ (see Fig.~\ref{fig:Anisotropyfig} in Appendix~\ref{sec: Anisotropy}). 

Typically, anisotropy increases as temperature is lowered causing domain walls to narrow and pin~\cite{TremoletdeLacheisserie2004, Hilzinger1972, Barbara1994}. The compound Dy$_3$Al$_2$ is one such example of this typical change in domain wall mobility due to anisotropy. As seen by a change in shape of the virgin curve, the domain walls in Dy$_3$Al$_2$ move freely at high $T$ while domain wall pinning occurs at low $T$~\cite{TremoletdeLacheisserie2004,Barbara1973,Zu2021Dy}. In contrast, the opposite is true in LaCrGe$_3$ which means a mechanism other than anisotropy is responsible for the domain wall behavior we observe.

We believe it is no coincidence that the local maximum of coercivity in the domain pinning region occurs at $72.5$\,K which is near the reported temperature of the crossover between FM1 and FM2. We will show that a change in the ferromagnetic state could be responsible for the domain wall pinning behavior in LaCrGe$_3$ in a following section. Next, we will demonstrate the relevance of the atypical domain wall behavior by showing how it can explain the previously observed, but not well understood, $M$\,v\,$T$ curves.

\subsection{\label{sec:SharkFin}Deconstructing the $M$\,v\,$T$ Anomaly}

\begin{figure}%[h]
\includegraphics[trim = 130 20 120 10]{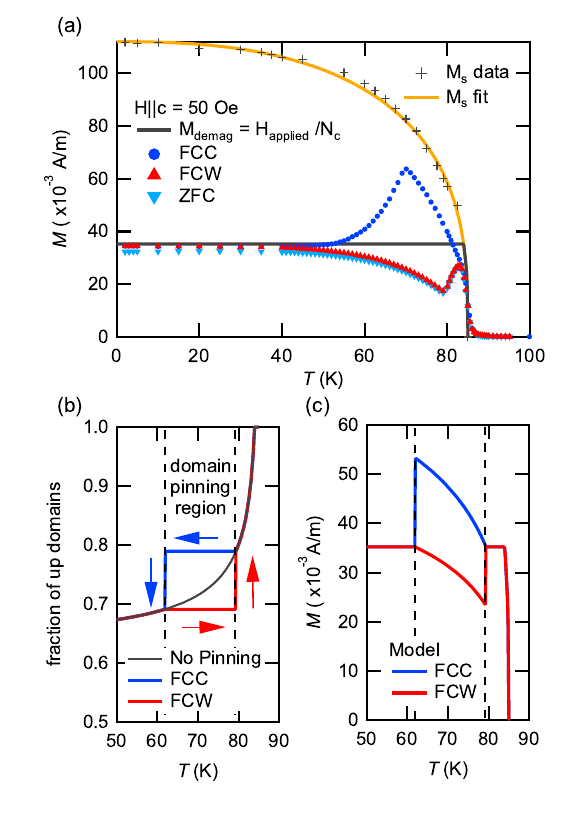}
\caption{\label{fig:MvT} (a) The non-standard field-cooled-cooling (FCC) and field-cooled-warming (FCW) magnetization data against $M$ determined by demagnetization theory (black curve), and the spontaneous magnetization (data shown as black crosses, fit by Kuz'min theory~\cite{Kuzmin2005} in yellow). (b) Without domain wall pinning, the domain fraction traces the black curve to satisfy demagnetization. The pinning region causes the domain fraction to trace different paths while cooling (blue curve) and warming (red curve). (c) By plotting the magnetization that results from these two different paths, we find that the previously unexplained features in both the FCC and FCW curves are faithfully reproduced. \modif{A direct comparison of the model and experimental data at $H = 50$\,Oe and at other applied fields is shown later in Fig.~\ref{fig:Appendix_MvT}b.}}
\end{figure}

The anomalous magnetization curve measured at low fields in LaCrGe$_3$ has previously been observed in studies of single crystals~\cite{Lin2013,XuArxiv} as well as polycrystals~\cite{BoschSantos2021a}. While the anomalous features in the curve were correctly attributed to changes in the magnetic domains and demagnetization effects~\cite{Lin2013}, in this section we will explain exactly how these features are caused by non-traditional domain behavior resulting from the crossover from FM1 to FM2.

To understand the unusual magnetization curve in LaCrGe$_3$, we first need to understand what we expect the magnetization curves to look like without domain wall pinning and depinning. In the absence of ferromagnetic domain formation, the magnetization should follow the spontaneous magnetization, $M_s$, represented by the yellow curve in Fig. \ref{fig:MvT}a. $M_s$ can indeed be measured if a sufficiently high field is applied ($H>NM_s$) to overcome domain formation, or equivalently, demagnetization effects. At lower applied fields, however, domains form and demagnetization theory can be used to determine the expected magnetization. When $H_{\textrm{applied}}<NM_{s}$, the expected magnetization is $M_{\textrm{demag}} = \frac{H_{\textrm{applied}}}{N}$, where $N$ is the demagnetization factor which is determined by the shape of the particular sample measured (see Appendix\,\ref{sec: Demag}). $M_{\textrm{demag}}$ is temperature independent and is depicted by the black line in Fig.~\ref{fig:MvT}a for $H = 50$\,Oe.

It is evident that the low field (below $NM_{s}$) FCC, FCW and ZFC data, shown respectively as dark blue, red, and light blue dots in Fig.~\ref{fig:MvT}a, deviate significantly from $M_s$ and $M_{\textrm{demag}}$ in the temperature region between $55$\,K and $82$\,K. In the FCC case, $M$ rises above $M_{\textrm{demag}}$ and reaches a distinct peak before decreasing and settling to a lower, temperature independent value matching $M_{\textrm{demag}}$. In ferromagnets, a decrease in the FCC magnetization along the easy axis is not trivial to explain. A spin-reorientation or an antiferromagnetic transition come to mind as possibilities, but neutron diffraction measurements have not found evidence for either~\cite{Cadogan2012, Gati2021}.

The FCW curve is also anomalous as it follows $M_{\textrm{demag}}$ at temperatures below $40$\,K, but then decreases below $M_{\textrm{demag}}$ before suddenly increasing to rejoin the FCC curve just below $T_\mathrm{C}$. It is unusual to have discrepancies between FCW and FCC magnetizations. Hysteresis between warming and cooling is unexpected, and is often seen when there is a first order phase transition between two magnetic states. In this case, however, there is no evidence for a phase transition between two different magnetic states in other probes~\cite{Taufour2016, Kaluarachchi2017, Gati2021}.

The ZFC curve is remarkably similar to the FCW curve, differing only by a small decrease in magnitude. ZFC curves often have a much smaller low $T$ magnetization than the FCC and FCW curves since ferromagnetic domains often have some degree of pinning at low temperatures and the field ($H = 50$\,Oe in this case) is only applied after cooling the sample to $T=2$\,K allowing domains to form. In LaCrGe$_3$, however, the domain walls at low temperatures move freely and therefore the ZFC curve nearly matches $M_{\textrm{demag}}$ at low temperatures.

Now, we will briefly describe the simple model that shows the strange FCC and FCW magnetization curves are the result of the pinning and depinning of domain walls discussed in the previous sections. First, we recognize that magnetization can be expressed in terms of the fraction of field aligned domains $n_{\textrm{up}}$. In the low-field ($H<NM_s$) regime we expect to measure $M_{\textrm{demag}}$, so $n_{\textrm{up}}$ is given by the equation~\cite{TremoletdeLacheisserie2004}

\begin{equation}
    n_{\textrm{up}} = \frac{1}{2M_{s}(T)}\textrm{min}\Big\{M_{\textrm{demag}},M_s\Big\} + \frac{1}{2}
\end{equation}
More details about this equation can be found in Appendix~\ref{sec:Modeling}. This equation yields the black line in Fig.~\ref{fig:MvT}b that shows the fraction of field-aligned magnetic domains as a function of temperature. Rather counterintuitively, the fraction of field-aligned domains decreases as temperature decreases, which happens in order to maintain the temperature independent $M_{\textrm{demag}}$ while $M_s$ increases as $T$ lowers.

The particular temperature at which the domain walls pin and de-pin is dependent on the applied field, the criteria for which is described in Appendix~\ref{sec:Modeling}. The domain fraction is held constant when the domains pin while depinning has the fraction return to the demagnetization value. As a result of the domain pinning, the domain fraction takes different paths when cooling and warming, seen as the blue and red lines in Fig.~\ref{fig:MvT}b, respectively. We can calculate the magnetization that results from these model FCC and FCW domain fraction paths with the following equation 

\begin{equation}
    M =  (2n_{\textrm{up}}-1)M_{s}(T)
\end{equation}

The resulting magnetization curves are presented in Fig.~\ref{fig:MvT}c. During field-cooled-cooling, there is no pinning immediately below $T_\mathrm{C}$ so $M$ takes the expected shape according to demagnetization theory. When the temperature is lowered into the pinning region, the domain fraction is held constant so $M$ increases proportionally to $M_s$. The downturn near $60$\,K is due to the domain walls depinning and the magnetization returning to $M_{\textrm{demag}}$. In the field-cooled-warming case, the deviation of $M$ from $M_{\textrm{demag}}$ begins around $60$\,K where $M$ begins to decease with increasing $T$. This decrease is due to the domain fraction being held constant by pinning so $M$ follows the shape of $M_s$. $M$ continues to decrease until around $80$\,K when the domain walls de-pin and consequently $M$ rapidly increases to rejoin the FCC curve. These model magnetization curves faithfully reproduce the characteristic features of the anomaly seen in the data. As expected, these features are smoothed out in the experimental data, where, unlike in our model, the pinning and depinning are partial and progressive. We find the match between the two, however, to be compelling evidence that the abnormal $M$\,v\,$T$ curve is due to domain wall pinning and depinning.

\subsection{\label{sec: Domain Pinning Evidence}How FM1 FM2 May Explain the Change in Domain Wall Mobility}

Here, we will motivate how the existence of FM1 and FM2 may explain the change in domain wall mobility in LaCrGe$_3$. We begin with the theory that describes how readily domain walls move in response to an externally applied magnetic field. Theory shows that in the absence of sample defects an applied magnetic field must overcome an energy barrier $\Delta \gamma$ to move a domain wall. It has been shown that $\Delta \gamma \propto e^{\frac{-\delta}{a}}$~\cite{TremoletdeLacheisserie2004, Hilzinger1972, Barbara1994}, where $\delta$ is the width of the domain wall and $a$ is the lattice spacing between local magnetic moments. 

In LaCrGe$_3$, the nearest neighbor Cr atoms lie along the $c$ axis and although XRD, neutron diffraction, and thermal expansion measurements~\cite{Gati2021} show that the $c$ lattice parameter unusually increases during cooling, the change is small and would increase pinning at low temperatures. As a result, whether domain walls are easy or difficult to move depends on how wide or how narrow they are, respectively. In a mean field, local moment model, the width of the domain wall $\delta$, is proportional to the spin of the magnetic ion $S$, the distance between magnetic moments $a$, the exchange constant $J_{ex}$ and the anisotropy constant $K$, as follows~\cite{Simon2013, TremoletdeLacheisserie2004, Blundell2001, Hilzinger1972}

\begin{equation} \label{eq:domainwidth}
    \delta = \pi S \sqrt{\frac{2J_{ex}}{aK}}
\end{equation}

The temperature dependence of $K$ that we measure agrees with theory (Appendix~\ref{sec: Anisotropy}) and would typically cause the domain walls to narrow and become pinned at lower temperatures. In contrast, we observe that domains walls in LaCrGe$_3$ become pinned at higher temperatures. \modif{Ruling out unusual anisotropy, we can consider the existence of two different exchange constants, $J_{ex1}$ and $J_{ex2}$, which lead to the two distinct anomalies at $T_{\textrm{C}}$ and $T_{x}$ that indicate the two different ferromagnetic states.} \modif{This is an appropriate consideration since having two different interactions is the basis for a `two-channel Stoner' model which was proposed to describe an itinerant d-electron ferromagnet such as LaCrGe$_3$, and predicts FM1 and FM2, as well as the appearance of an AFM phase under pressure~\cite{Wysokinski2019SR}. In addition, this `two-channel Stoner' model is compatible with UGe$_2$~\cite{Hardy2009PRB} and ZrZn$_2$~\cite{Kimura2004} where the existence of FM1 and FM2 are well established.}

%Ruling out unusual anisotropy, we can consider changes in the exchange constant, $J_{ex}$, as potentially responsible for the unusual changes in domain wall mobility. While $J_{ex}$ is usually constant throughout a single ferromagnetic state, the existence of two ferromagnetic states allows for a change in $J_{ex}$ and we can check to see if the change could be responsible for the domain pinning in FM1 and the depinning in FM2. 

We start by taking the ratio of the domain wall widths in the two FM states and by realizing that $J_{ex}$ is related to $T_\mathrm{C}$ \cite{Blundell2001}

 \begin{equation} \label{Eq:dd4}
    \frac{\delta_{FM1}}{\delta_{FM2}} = \sqrt{\frac{T_{C}}{T_{x}}} \frac{\sqrt{\frac{S_1^2}{S_1(S_1+1)}}}{\sqrt{\frac{S_2^2}{S_2(S_2+1)}}} 
\end{equation}

\modif{$T_{\textrm{C}} = 85$\,K} and experimental data at ambient pressure shows that the crossover occurs at \modif{$T_x \approx 70$\,K} \cite{Kaluarachchi2017} which nearly corresponds to the coercivity maximum discovered in this work, further suggesting the crossover influences the depinning of domain walls. Since we cannot directly measure $M_s(0)$ in FM1, we instead estimate the change in spin between FM1 and FM2 by the loss of magnetic scattering between the two phases as reported previously~\cite{Taufour2018}. Early theories of the resistivity of a ferromagnetic metal~\cite{PeskiTinbergen1963, Kasuya1956, Gennes1958} relate the spin of the local magnetic moments to the resistivity within and outside of the ferromagnetic state. The result is: 

\begin{equation}\label{DW_Theory_Result}
    \frac{\delta_{FM1}}{\delta_{FM2}} \approx 0.56
\end{equation}

Our rough theoretical approximation shows that the domain walls in the high temperature FM1 state are expected to be shorter, and therefore more difficult to move, than the domain walls in the lower temperature FM2 state. This result from allowing for two ferromagnetic states is consistent with our experimental observation that the magnetic domains are pinned at high temperatures near $70$\,K and subsequently de-pin as the temperature is lowered, resulting in an unusual magnetization curve and temperature dependence of coercivity.

Applying a similar approximation to the expression for the energy cost per unit area of a Bloch domain wall~\cite{Blundell2001},
\begin{equation}
    \sigma_{BW} = \pi S \sqrt{\frac{2J_{ex}K}{a}}
\end{equation}
yields the same ratio as Eq.~\ref{DW_Theory_Result}, which implies that the cost of a domain wall is larger in the FM2 state. This result means that we expect to see fewer domain walls in the FM2 phase compared to the FM1 phase, which is consistent with our MOKE images presented in Fig.~\ref{fig:MOKE} that show the domains are larger in the FM2 phase than in the FM1 phase. 

\section{Conclusion}
In conclusion, we recognize features in spatially resolved images of magnetic domains, as well as features in bulk magnetization that show a unique change in domain wall mobility in LaCrGe$_3$. We then showed the relevance of the domain wall pinning at high temperatures followed by depinning at low temperatures by using it to explain anomalous features in the $M$\,v\,$T$ curves that were previously observed, but not well understood. Finally, we find that a crossover between FM1 and FM2 could cause the change of domain wall mobility we observe. Our discovery joins a number of other probes that show anomalies in a similar temperature region and support the existence of FM1 and FM2 in LaCrGe$_3$. Our work can inspire similar magnetic domain analysis on other systems to probe crossovers between multiple ferromagnetic states, which are uncannily common in fragile FM systems.

\begin{acknowledgments}
V.T. and R.R.U. acknowledge support from UC Davis Startup funds, the UC Laboratory Fees Research Program (LFR-20-653926) and the Physics Liquid Helium Laboratory Fund. X.D.Z acknowledges support from Fudan University in the
form of a Visiting Lecture Professorship.
\end{acknowledgments}

\appendix

\section{\label{sec:experimentaldetails}Experimental Method Details}

\subsection{\label{subsec:synthesis}Crystal Synthesis}
 We use a non-stoichiometric ratio of 12.75:12.75:74.5 of Ames Laboratory La pieces, 4N Aesar Cr crystallites, and 6N Alfa Aesar Puratronic Ge plates. The constituent elements were arc melted together before being placed into a Canfield Crucible Set~\cite{Canfield2016}. 

The crucible set is sealed in a quartz tube in a 160\,mmHg partial pressure of Ar. The sealed ampoule is heated up from room temperature to 1100$^\circ$C over a 4 hour period and held at that temperature for 5 hours to ensure complete dissolution. The reaction is cooled to 1000$^\circ$C in 2 hours before being slowly cooled to 850$^\circ$C over an 83 hour period. At 850$^\circ$C, the ampoule is quickly removed from the furnace and placed into a centrifuge where the liquid flux is separated from the solidified crystals. LaCrGe$_3$ single crystals form hexagonal prisms where the $a$ and $b$ axes form the hexagonal faces that grow along the $c$ axis.

We found that reactions placed closer to the door of the furnace yielded large single crystals while those placed deeper into the furnace resulted in delicate needles. This observation leads us to believe that a thermal gradient is beneficial to this particular synthesis. 

A selection of crystals were crushed into a fine powder and powder x-ray diffraction data was collected with a Rigaku MiniFlex diffractometer. We confirmed that these crystals were of the correct phase by comparing the measured pattern to previous reports~\cite{Bie2007}.

\subsection{\label{subsec:AppendixMagDetails}Magnetization Measurements}
We use three different measuring procedures to measure magnetization as a function of temperature ($M$\,v\,$T$). Field-cooled-cooling (FCC), field-cooled-warming (FCW) and zero-field-cooling (ZFC). For FCC, the field is applied at $T = 300$\,K and the magnetization of the sample is measured while temperature is lowered to base temperature, $T = 2$\,K. With the field still on, the magnetization is measured while increasing temperature up to $T = 300$\,K for the FCW measurement. For ZFC, the sample is cooled from above $T_\mathrm{C}$ to base temperature in zero applied field. Then the field is applied and the magnetization is measured while the sample is warmed.

Hysteresis loops were measured starting in a zero-field-cooled state to observe the virgin magnetization curve before sweeping field up to $\pm$7\,T. Before each measurement, the sample temperature was raised above $T_\mathrm{C}$ to 110\,K in order to clear the magnetic history of the sample before the subsequent measurement. Furthermore, the field was systematically oscillated to zero to minimize the remnant field in the magnet and the sample chamber. Our calibration with a paramagnetic standard shows that the remanent field of the magnet after this procedure is roughly $8$\,Oe. These steps are important because we are specifically interested in the coercive field and the virgin curve which are relatively low-field phenomena.

\section{Additional MOKE Images}

\begin{figure*}
\centering
\includegraphics[trim = 10 30 0 0]{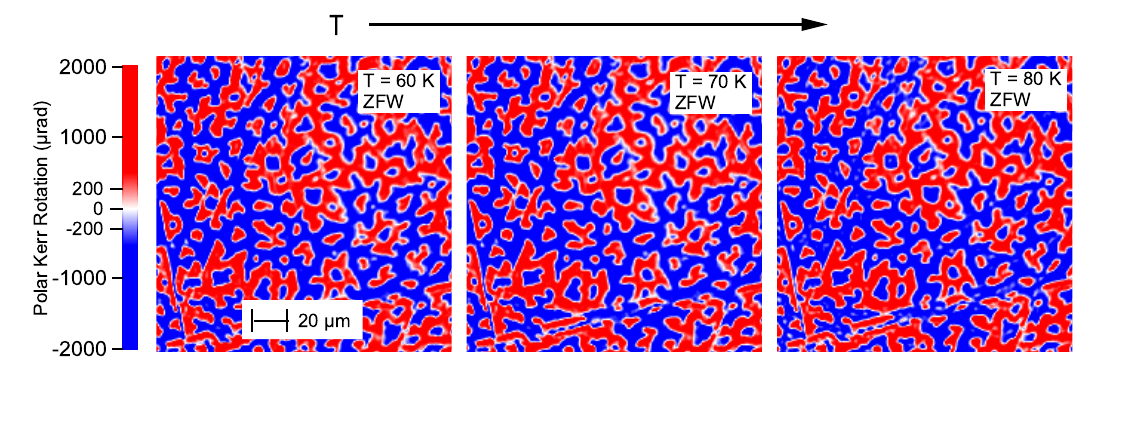} %trim is left bottom right top
\caption{\label{fig:MOKE_ZFW} Polar Kerr rotation images of the same polished $ab$ face in a similar area as Fig.~\ref{fig:MOKE}. These images were taken while warming in zero field from $11.5$\,K after initially cooling in an applied field of $H=3000$\,Oe. Since the domain walls move freely at low temperatures, once the field is turned off at $11.5$\,K, there is no detectable remanent magnetization. Unlike the ZFC case shown in Fig.~\ref{fig:MOKE}, there is no change in the domain structure when warming from FM2 to FM1. This lack of change, however, is consistent with a domain wall pinning region in FM1.}
\end{figure*}

\begin{figure*}
\centering
\includegraphics[width=\textwidth]{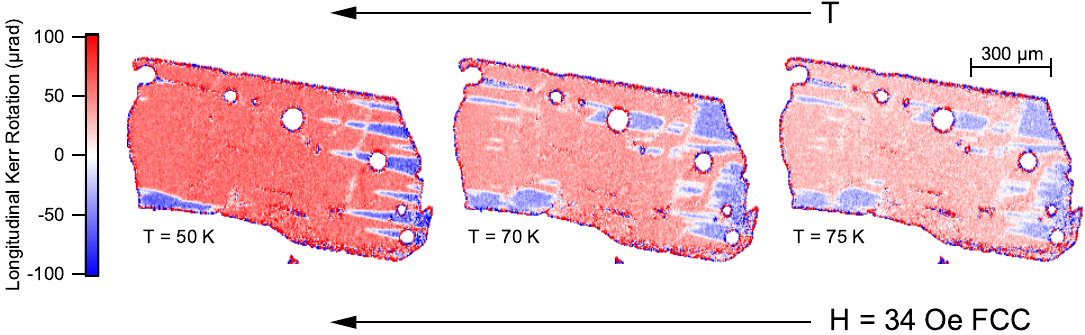} %trim is left bottom right top
\caption{\label{fig:MOKE_ACPlane} Longitudinal Kerr rotation images of an as-grown $ac$ face of LaCrGe$_3$ acquired while field-cooling in $H=34$\,Oe applied to the left as indicated by the arrow. \modif{Positive values of Kerr rotation are shown in red and represent regions where $M$ is aligned with the applied field. Negative values are shown in blue and indicate regions where $M$ is anti-aligned with the field.} We observe similar domain behavior to that of the $ab$ plane discussed in the main article. There is little change in the domain structure between $75$\,K \modif{and} $70$\,K, which we attribute to domain pinning in the FM1 state. Below $60$\,K, there is a significant change in the domain structure, as shown by the image taken at $50$\,K, which we attribute to the depinning of domain walls in the FM2 state.}
\end{figure*}

While the MOKE images shown in Fig.~\ref{fig:MOKE} clearly demonstrate a change in domain structure on either side of the FM1-FM2 crossover, in this section, we provide additional MOKE studies supporting our observation of domain pinning in FM1 and depinning in FM2.

Fig.~\ref{fig:MOKE_ZFW} shows images taken while warming in zero field (ZFW) from $11.5$\,K from a similar area as Fig.~\ref{fig:MOKE}. While the sample was cooled to $11.5$\,K in $H_{\textrm{applied}}=3000$\,Oe, removing the field at low temperatures resulted in a net zero magnetization, which matches demagnetization theory and is consistent with the lack of domain wall pinning at low temperatures. Upon warming in zero field, the images taken throughout the temperature range below $T_\mathrm{C}$ are indistinguishable to the eye. Unlike the ZFC images shown in Fig.~\ref{fig:MOKE}, these ZFW images do not show a considerable change in the domain structure between FM1 and FM2. This result is consistent with domain wall pinning at high temperatures and is not in conflict with the existence of the FM1 state.

We also revisited the same $ac$ plane studied in~\cite{Zhu2021RSI}, this time using liquid helium to obtain images below $77.4$\,K. These images, shown in Fig.~\ref{fig:MOKE_ACPlane}, were taken while cooling in an applied field of $H_{\textrm{applied}}=34$\,Oe in the direction indicated by the arrow in the figure. The images taken at $75$\,K and $70$\,K are representative of those in the FM1 state showing little change in domain structure. Below $60$\,K, the domain structure changes significantly as shown by the $50$\,K image.

\section{\label{sec:Modeling}Modeling the $M$\,v\,$T$ Curve}
\begin{figure}%[h]
\includegraphics[trim = 125 20 120 0]{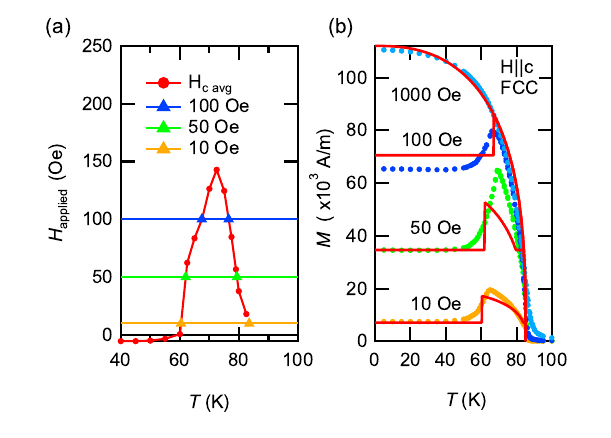}
\caption{\label{fig:Appendix_MvT} (a) By finding the intersection between $H_{c\textrm{ avg}}$ and a particular applied field, we find the temperatures at which the domains pin and de-pin for each field.  (b) We find that this criteria is in good agreement with the data, particularly at lower applied fields.}
\end{figure}

In this section, we will develop a simple model involving the spontaneous magnetization, demagnetization factors, and ferromagnetic domains to show how the unusual features in the $M$\,v\,$T$ curve can be explained by the pinning and depinning of domain walls. 

In experiment, the spontaneous magnetization, $M_s(T)$, is a fingerprint for a ferromagnetic compound. Experimentally, $M_s$ is found from $M$\,v\,$H$ measurements along the easy-axis by taking the y-intercept of the line fit to the saturated region at high fields. This procedure can be performed at various temperatures up to $T_\mathrm{C}$ (where $M_s = 0$) to construct $M_s(T)$. In a strictly local moment ferromagnet, $M_s(T)$ is theoretically the maximum magnetization one would measure while measuring $M$\,v\,$T$. We do not observe any unusual behavior in $M_s(T)$ and it fits well according to the theory by Kuz'min~\cite{Kuzmin2005}, as seen in Fig.~\ref{fig:MvT}a. 

In any finite sample with a magnetization $M$, there is a demagnetization field $H_d = -NM$, where $N$ is a demagnetization factor determined by the sample dimensions (discussed further in Appendix \ref{sec: Demag}). As a result, the H-field inside the sample, $H_{\textrm{int}}$ = $H_\textrm{applied} + H_d$, is less than the applied field. In ferromagnets with an easy magnetization direction and where domain wall pinning is not present, the consequence of $H_d$ is readily seen when measuring $M$\,v\,$H$ from a zero-field-cooled state (where $M\approx0$ when $H=0$). In these cases, the initial magnetization increases linearly with the applied field with slope of $1/N$ until it reaches the spontaneous magnetization $M_s$ at field $|H_{d,max}| = NM_s$. In the linearly increasing region ($H_{\textrm{applied}}<|H_{d,max}|$),

\begin{equation}\label{eq.magdemag}
    M = \frac{1}{N}H_{\textrm{applied}}
\end{equation}
which tells us that the magnetization \emph{only} depends on the sample shape via the demagnetization factor, $N$. Notice that this magnetization is temperature independent, and we can see that Eq.~\ref{eq.magdemag} faithfully reproduces the magnetization below $40$\,K at $50$\,Oe as seen in Fig.~\ref{fig:MvT}a. 

From a single rotating moment picture and considering that $H_{\textrm{int}} = 0$ when $H_{\textrm{applied}}<|H_{d,max}|$, it makes sense that we do not immediately measure $M_s$, as there is no field to align the single rotating moment along our axis of measurement. Alternatively, and \emph{equivalently}, we can consider the formation of ferromagnetic domains and the motion of the domain walls between them. From the domain point of view, the linearly increasing region of $M$\,v\,$H$ is due to the applied field moving ferromagnetic domains walls, and adjusting the fraction of aligned and anti-aligned domains in the sample. In a material with strong uniaxial anisotropy, such as LaCrGe$_3$, we only need to take into account the fraction of domains aligned with the field, $n_{\textrm{up}}$, and those anti-aligned with the field, $n_{\textrm{down}}$. Since $n_{\textrm{up}} + n_{\textrm{down}}=1$ and since the maximum magnetization is $M_{s}$, 

\begin{equation}
    M = n_{\textrm{up}}M_{s}-n_{\textrm{down}}M_{s}
\end{equation}

\begin{equation}\label{eq.magdomains}
    M =  (2n_{\textrm{up}}-1)M_{s}
\end{equation}

In the low field ($H_{\textrm{applied}}<NM_s$) region, the demagnetization picture, Eq.~\ref{eq.magdemag}, and the domain picture, Eq.~\ref{eq.magdomains} are both valid and we can equate them to find the temperature dependence of $n_{\textrm{up}}$~\cite{TremoletdeLacheisserie2004}. The result is

\begin{equation}\label{eq.domainfraction}
    n_{\textrm{up}} = \frac{1}{2M_{s}(T)}\textrm{min}\Big\{\frac{H_{\textrm{applied}}}{N},M_s\Big\} + \frac{1}{2}
\end{equation}
where the minimum function is used to correctly model $M$ just below $T_\mathrm{C}$ and avoid the unphysical result where the measured magnetization is greater than $M_s$. This equation allows us to plot the fraction of domains as a function of temperature. When the domain walls are free to move, in order to fulfill $M=1/N$, the fraction of up domains must decrease as temperature decreases, as seen by the black line in Fig.~\ref{fig:MvT}b. This result is counterintuitive when we think of the typical magnetization curve which does not consider demagnetization effects and therefore does not exhibit the decrease in the fraction of field-aligned domains. If we instead hold the domain fraction, $n_{\textrm{up}}$, constant for a range of temperature, we recover the unusual features observed in the experimental $M$\,v\,$T$ curve for both FCC and FCW measuring methods (Fig.~\ref{fig:MvT}c).

In Fig.~\ref{fig:Appendix_MvT}, we show the criteria used to choose the temperatures at which the domains pin and de-pin in our model. We developed $H_{c \textrm{ avg}} = (H_{c} + H_{\textrm{shoulder}})/2$ to take into account the wide range of field the domain walls move in the high $T$ hysteresis loops. In this equation, $H_c$ is the typical definition of the coercive field, the field required to make $M=0$ after it was previously saturated to $M=M_s$. We define $H_{\textrm{shoulder}}$, as the field at which the magnetization just begins to change from its fully polarized value. In the case of the rectangular hysteresis loops at low temperature, these two measures of the coercivity are the same, i.e. $H_c = H_{\textrm{shoulder}}$. As evident from the virgin magnetization curves, as well as the low-field $M$\,v\,$T$ data, however, domain wall pinning is not responsible for the low $T$ region of coercivity.

In conclusion, we recognize that a textbook equation relating demagnetization effects and magnetic domains can be used to model low-field ferromagnetic magnetization curves. With Eq.~\ref{eq.domainfraction} in an experimentalist's toolbox, it is possible to check whether anomalous features in $M$\,v\,$T$ are due to demagnetization or the unusual behavior of magnetic domains before more exotic possibilities are suggested or explored.

\section{\label{sec: Demag}Determining Demagnetization Factors}

\begin{figure}
\includegraphics[width=0.8\columnwidth]{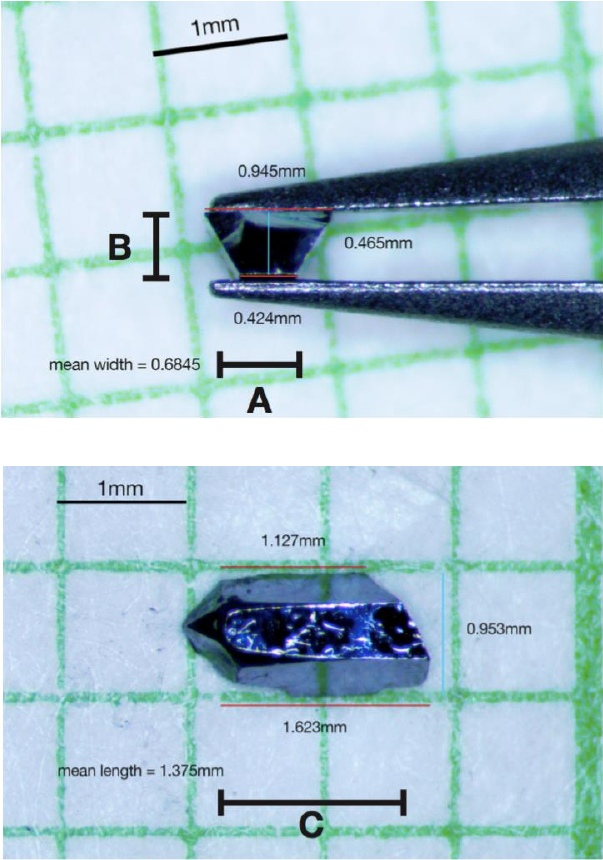}
\caption{\label{fig:Crystal} Most of the magnetic measurements were done on the LaCrGe$_3$ crystal pictured. One side was lightly polished flat and parallel to the opposite face. We determined the size from pictures of the sample on mm grid paper. From these measurements we were able to estimate the demagnetization factors $N_a$, $N_b$ and $N_c$.}
\end{figure}

\begin{figure}
\includegraphics[trim = 120 10 110 0]{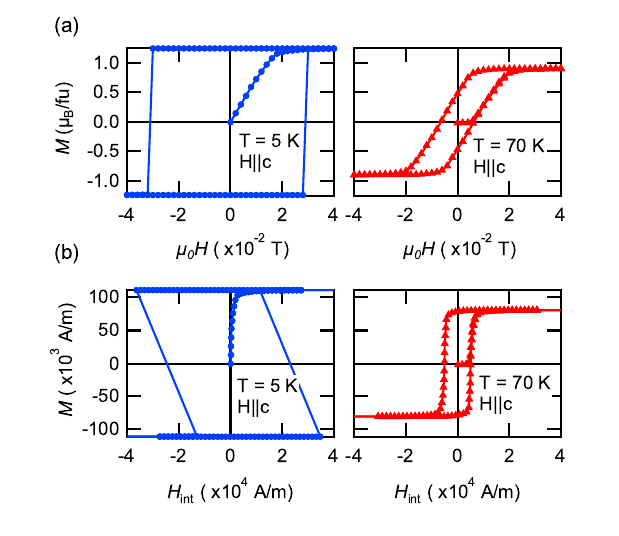}
\caption{\label{fig:demagcorrect}(a) $M$\,v\,$H$ curves and (b) the demagnetization corrected $M$\,v\,$H_{\textrm{int}}$ curves. By taking into account the demagnetization field resulting from the finite size of our magnetized sample, we can plot magnetization against the internal field, $H_{\textrm{int}}$.}
\end{figure}

Since our study involves magnetic domain behavior, the demagnetizing field, $H_d = -NM$, plays an important role in our analysis. While the measured magnetization, $M$, is a property of the compound in the particular temperature and field environment, the demagnetization factor $N$ is unique to the size and shape of the sample. In this section, we will explain how we determined N for the sample we measured in this article.

Pictures of the sample used for all of the measurements which include $N$ in the analysis are shown in Fig.~\ref{fig:Crystal}. We measured the dimensions of the sample from these photographs taken against mm-grid paper. We then used the mean length and mean width to approximate the half-hexagonal rod as a rectangular prism, so we could use the formula given by Aharoni~\cite{Aharoni1998} to estimate the demagnetization factors along each axis. With this method, we found $N_a = 0.341$, $N_b = 0.492$, and $N_c = 0.166$. 

$N_c$ can also be determined from $M$\,v\,$H$ measurements in temperature ranges where there is no domain wall pinning and the easy axis virgin magnetization forms a straight line with slope \modif{$1/N_c$} through the origin. In LaCrGe$_3$, this condition is met for temperatures $\sim50$\,K and below. For temperatures around $60$\,K and above, however, the virgin curves do not initially follow demagnetization theory due to the presence of domain wall pinning. Experimentally, there is a small variation of the extracted $N$ across virgin magnetization curves measured at different temperatures, so in practice, the lowest value found is chosen to avoid the unphysical consequence that $H_{\textrm{int}}$ is negative. From our magnetization measurements, we found $N_c = 0.1129$. A difference of $\sim0.05$ for $N_c$ is within reason according to Lamichhane et al.~\cite{Lamichhane2016b} where a similar method for finding $N$ from sample size followed by correcting $N$ from magnetization measurements was used.

This value of $N_c$ is reasonable for two reasons. First, we see that $M = \frac{H_\textrm{applied}}{N_c}$ very closely matches the measured $M$ in the temperature independent region of the $M$\,v\,$T$ curve shown in Fig.~\ref{fig:MvT}a that we argue is simply the demagnetization value. Second, as shown in Fig.~\ref{fig:demagcorrect}b, using $N_c$ to compute $H_{\textrm{int}}$ makes the virgin curve increase nearly vertically to saturation, as expected for an easy axis ferromagnet.

With a working value of $N_c$, $N_a = 0.3631$ and $N_b = 0.524$ are found from ratios determined from the sample dimensions. For $H||ab$ measurements, the field is applied along the $A$ dimension labelled in Fig. \ref{fig:Crystal}a, so $N_a$ is used to calculate $H_{\textrm{int}}$, which is necessary for calculating anisotropy constants.

\modif{\section{\label{sec: SampleDependence}Sample Dependent Features}}

\begin{figure}[t]
\includegraphics[width = \columnwidth]{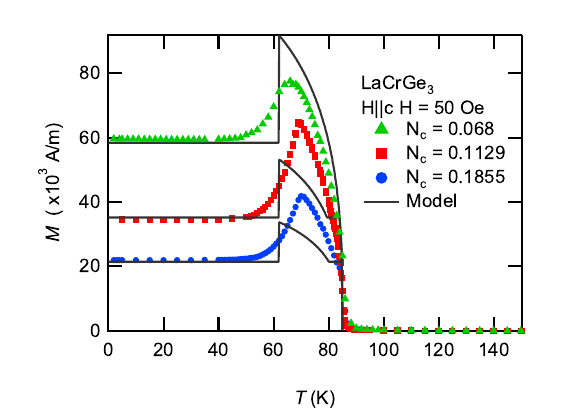}
\caption{\label{fig:SD_MvT} \modif{A comparison of the magnetization as a function of temperature at $H=50$\,Oe for three different samples. When $T<50$\,K, $M=H_{\textrm{applied}}/N_c$ as dictated by demagnetization theory. The demagnetization factor along the $c$ axis, $N_c$, is sample dependent and as a result, longer samples with smaller $N_c$ have larger $M$ for a given applied field in this low temperature region. The pinning and subsequent depinning of magnetic domains below $T_c$ is observed in all samples measured.}}
\end{figure}

\modif{Any discussion of magnetic domains or hysteresis requires us to consider variations from sample to sample. In this section, we repeat many of the magnetic measurements we performed in the main article on two additional single crystals with different shapes and sizes in order to help explain which magnetic features are intrinsic to LaCrGe$_3$ and which features are sample-dependent. Of the characteristics that are sample-dependent, we determine which ones are due to differences in demagnetization factor, and which properties may instead be due to defects in the sample.}

\modif{In LaCrGe$_3$ measured along the easy $c$ axis, the effects of $N_c$ are apparent in both $M$\,v\,$T$ and $M$\,v\,$H$ measurements. The unusual $M$\,v\,$T$ curve is only measured at low fields, specifically when $H_{\textrm{applied}} < N_cM_s$. According to this relationship, we expect short samples with large $N_c$ to show the anomaly up to higher fields than longer samples with smaller $N_c$. We also note that the temperature-independent magnetization observed when $T<50$\,K is exactly dictated by demagnetization theory, $M = H_{\textrm{applied}}/N_c$. As a result, for a given applied field, a longer sample will settle to a higher magnetization compared to a shorter sample. This can be seen in Fig.~\ref{fig:SD_MvT}, where we plot $M$\,v\,$T$ at $H=50$\,Oe for three samples, along with our model which accounts for the different values of $N_c$.}

\begin{figure}
\includegraphics[width = \columnwidth]{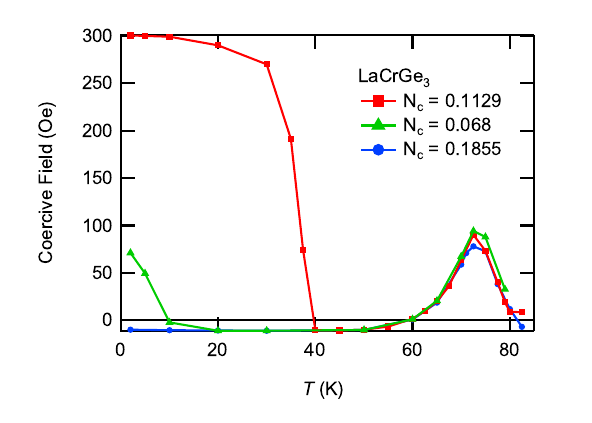}
\caption{\label{fig:SD_Coercivity} \modif{A comparison of the temperature dependence of the coercive field, $H_c$, across three different samples. Below $40$\,K, the hysteresis loops vary significantly across the three samples. In one sample, no coercivity was observed. Furthermore, there is no trend based on sample size; longer samples do not necessarily have higher coercive fields, as shown by comparing the green curve for the $N_c = 0.068$ sample to the red curve for the shorter-in-length $N_c = 0.1129$ sample. This suggests that the low temperature hysteresis is due to defects, but more research is needed to confirm. On the other hand, the coercive field is nearly identical across all three samples when it reappears at high temperatures ($60$\,K$<T<T_\textrm{C}$). We take this as evidence that the hysteresis in this range of temperatures is due to an intrinsic property of LaCrGe$_3$.}}
\end{figure}

\modif{The initial magnetization curve is a signature of $N_c$ in $M\,$v\,$H$ and was discussed in Appendix~\ref{sec: Demag}. We find that our additional samples also have virgin curves that follow demagnetization theory when $T\lesssim60$\,K. In addition, we recognize that when $M$ starts to increase in the high temperature ($T\gtrsim60$\,K) hysteresis loops, it does so with a slope equal to $1/N_c$. Interestingly, the slope is also $1/N_c$ when the magnetization reverses from $\pm M_s$ to $\mp M_s$, which suggests that in this high temperature range, the same underlying mechanism is responsible for both the magnetization reversal and the initial increase in $M$ up to $M_s$~\cite{TremoletdeLacheisserie2004}. Finally, since our samples have a similar $H_c$ in this range of temperatures, as shown in Fig.~\ref{fig:SD_Coercivity}, when we factor in $N_c$ to plot $M$\,v\,$H_{\textrm{int}}$, we find that the hysteresis loops in this high temperature region are nearly identical across our three samples.}

\modif{The hysteresis loops at low temperature, however, are not identical across all samples. Fig.~\ref{fig:SD_Coercivity} shows that at low temperatures, $H_c$, as well as the temperature at which $H_c$ disappears, are sample-dependent. The shortest sample measured ($N_c = 0.1855$) has a negligible $H_c$, while the longest sample ($N_c = 0.068$) has less than one third the $H_c$ as the sample measured in the main text ($N_c = 0.1129$), and this coercivity disappears at a relatively low temperature. It is clear that the low-temperature hysteresis loops are wildly variable, with no obvious trend based on sample size. Naively, we might expect longer samples to have larger $H_c$ due to the increased energy cost of forming the longer domain wall necessary to nucleate a reversed domain. This hypothesis, however, does not match the $H_c$\,v\,$T$ data across the three samples presented here, along with the additional two from Xu et al.~\cite{XuArxiv}. Therefore, unlike the magnetic features discussed previously where the sample-dependence was determined by the size of the sample via the demagnetization factor, we are unable to draw conclusions regarding the relationship between $N_c$ and $H_c$ at low temperatures. Exploring the role of sample defects may help explain the sample-dependence of these low temperature hysteresis loops.}\\

\section{\label{sec: Anisotropy}Calculating Anisotropy Constants}

\begin{figure}[!htb]
\includegraphics[trim = 110 10 110 0]{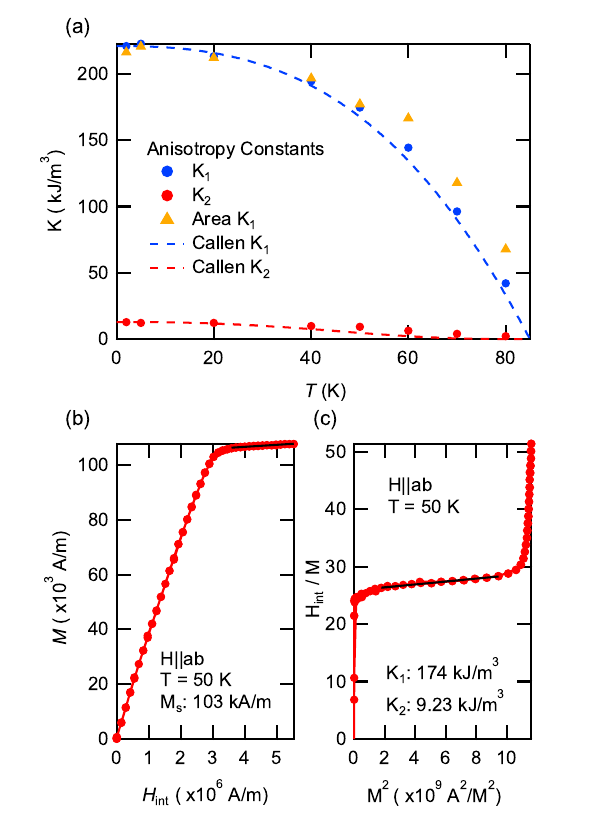} %trim is left bottom right top
\caption{\label{fig:Anisotropyfig} (a) The anisotropy constants as a function of temperature. \modif{There do not appear to be any anomalies in $K_1$ or $K_2$, which may initially be surprising given the strange behavior in $M$\,v\,$T$.} (b) $M$\,v\,$H$ measured along the hard axis. The linear region at high field is fit to a line to extract the spontaneous magnetization $M_s$. (c) By plotting $H_{\textrm{int}}$\,v\,$M^2$ and fitting the linear portion to Eq.~\ref{eq:k1k2}, we can extract the anisotropy constants $K_1$ and $K_2$.}
\end{figure}

In this section, we summarize three different ways to calculate the anisotropy constants, $K_1$ and $K_2$, for a ferromagnet with a hard and easy axis, such as LaCrGe$_3$. The results of our anisotropy analysis are summarized in Fig.~\ref{fig:Anisotropyfig}a, and we find that $K_1$ and $K_2$ in LaCrGe$_3$ follow a standard temperature dependence. So unlike in the case of Gd, anisotropy alone cannot explain the unusual temperature dependence of $H_c$ in LaCrGe$_3$.

The first way to measure the anisotropy constants is a method first detailed by Sucksmith~\cite{Sucksmith1954}. The Sucksmith method involves measuring $M$ as a function of $H_\textrm{applied}$ along the hard axis of the ferromagnet, and relating that measurement to an expression for the minimum of an energy density. The energy density we consider is a sum of the anisotropy energy density and the Zeeman energy.

\begin{equation}
    \epsilon = \epsilon_{\textrm{anisotropy}} + \epsilon_{\textrm{Zeeman}}
\end{equation}

The anisotropy energy density is given by
\begin{equation}
    \epsilon_{\textrm{anisotropy}} = K_1\sin^2\theta + K_2\sin^4\theta
\end{equation}
where $K_1$ and $K_2$ are the anisotropy constants and $\theta$ is the angle between magnetic moment, $M$, and the easy magnetic axis direction (the $c$ axis for LaCrGe$_3$).

The Zeeman energy, which describes the interaction between the moment and the applied magnetic field, is given by
\begin{equation}
    \epsilon_{\textrm{Zeeman}} = -\vec{M}\cdot \vec{B} = -M\mu_0H\cos{\phi}
\end{equation}
where $\phi$ is the angle between the moment and the applied magnetic field $H$. For this energy to be correct, we need to consider the demagnetization field and instead of $H$ we should explicitly use $H_{\textrm{int}}$

\begin{equation}
    H_{\textrm{int}} = H_{\textrm{applied}} + H_{\textrm{demag}} = H_{\textrm{applied}} - NM_{\textrm{measured}}
\end{equation}
where $M_{\textrm{measured}}$ is the measured sample magnetization.

When measuring the magnetization along the hard axis, the measured magnetization is $M_{\textrm{measured}}= M_s\sin{\theta}$ in which $M_s$ is the spontaneous magnetization. $M_s$ is given by the y-intercept of the line fit to the saturated magnetization at high field as depicted in Fig.~\ref{fig:Anisotropyfig}b. As a result, the theta-dependent energy density is

\begin{equation}
    \epsilon = K_1\sin^2{\theta} + K_2\sin^4{\theta} - \mu_0M_s H_{\textrm{int}}\sin{\theta}
\end{equation}
which can be minimized with respect to $\theta$ by taking the derivative and setting it equal to zero as follows
\begin{equation}
    \frac{\delta\epsilon}{\delta\theta} = 2K_1\sin{\theta}\cos{\theta + 4K_2\sin^3{\theta}\cos{\theta}-\mu_0M_sH_{\textrm{int}}\cos{\theta}} = 0
\end{equation}
In this equation, $\cos{\theta} = 0$ or $\theta = \pi$ is simply the trivial solution in which the magnetization is all along the hard axis. So we are left with

\begin{equation}
    2K_1\sin{\theta} + 4K_2\sin^3{\theta}-\mu_0M_sH_{\textrm{int}} = 0
\end{equation}
Remembering that $M_{\textrm{measured}} = M_s\sin{\theta}$, we can rewrite the $\sin{\theta}$ as $\frac{M}{M_s}$. With some further arrangement, we have

\begin{equation}\label{eq:k1k2}
    \frac{\mu_0H_{\textrm{int}}}{M} = 2K_1\frac{1}{M_s^2} + 4K_2\frac{1}{M_s^4}M^2
\end{equation}

Finally, by plotting $\frac{\mu_0H_{\textrm{int}}}{M}$\,v\,$M^2$ we find $K_2$ from the the slope which is $\frac{4}{M_s^4}K_2$ and $K_1$ from the intercept, $\frac{2}{M_s^2}K_1$, as seen in Fig.~\ref{fig:Anisotropyfig}c. To get the temperature dependence of $K_1$ and $K_2$, the Sucksmith method is performed on hard axis $M$\,v\,$H$ measurements at various temperatures below $T_\mathrm{C}$. Our results for $K_1$ and $K_2$ are shown, respectively, as the blue and red dots in Fig.~\ref{fig:Anisotropyfig}a.
 
We can compare our measured anisotropy to its theoretical temperature dependence given by the Callen-and-Callen law~\cite{Lamichhane2020, Callen1960}. Given $K_1^0$ and $K_2^0$, the anisotropy constants at zero temperature, and $M_s(T)$, the temperature dependence of the spontaneous magnetization, the anisotropy constants at temperature $T$ are given by 

\begin{equation}
    K_1(T) = \Big(K_1^0 + \frac{7}{8}K_2^0\Big)\Big(\frac{M_s(T)}{M_s(0)}\Big)^{3} - \frac{7}{8}K_2^0\Big(\frac{M_s(T)}{M_s(0)}\Big)^{10}
\end{equation}

\begin{equation}
    K_2(T) = K_2^0\Big(\frac{M_s(T)}{M_s(0)}\Big)^{10}
\end{equation}

For this Callen-and-Callen analysis, we used the $T = 2$\,K values from the Sucksmith method as $K_1^0$ and $K_2^0$. For $M_s(0)$, we use the $T = 2$\,K value of the spontaneous magnetization which is calculated from the y-intercept of the line fit to the saturated portion of an $M$\,v\,$H$ measurement with $H||c$. $M_s(T)$ is calculated from $M_s(0)$ using theory from Kuz'min~\cite{Kuzmin2005}. The theoretical $K_1(T)$ and $K_2(T)$ from this Callen-and-Callen analysis are shown, respectively, as the dashed blue and red lines in Fig.~\ref{fig:Anisotropyfig}a. Our measured values derived from the Sucksmith analysis are in agreement with the Callen-and-Callen law throughout the temperature range.

The final method we used to extract anisotropy from magnetization data involves computing the area between the easy and hard axis magnetization curves in $M$\,v\,$H$~\cite{Bozorth1937}. With this area method, $K_1$ is given by

\begin{equation}
    K_1 = \epsilon_{001} - \epsilon_{100}  
\end{equation}
where $\epsilon_{001}$ and $\epsilon_{100}$ are the magnetic energies along the easy axis ($c$ axis) and the hard plane ($ab$ plane) respectively. The energy densities are calculated by the integral
\begin{equation}
    \epsilon = \mu_0\int_{0}^{H_{sat}}MdH = \mu_0\int_{0}^{M_{sat}}HdM
\end{equation}
The results of using the area method are shown as the golden triangles in Fig.~\ref{fig:Anisotropyfig}a. The values for $K_1$ agree at temperatures below $50$\,K, however, they appear to systematically deviate at higher temperatures. We can understand this deviation by looking at the $M$\,v\,$H$ data at these temperatures which show a discrepancy in the saturation magnetization between the two axes. We have found that when measuring along the hard axis of a ferromagnetic sample, small sample displacements from the axial center of the sample chamber can cause large deviations in the measured $M$ at high fields. So despite our best efforts to axially center the sample in the MPMS, we found small discrepancies between easy axis and hard axis saturation magnetizations. Since the area analysis involves both easy and hard axis data, it is particularly sensitive to small mis-alignments compared to the Sucksmith method which only involves hard axis data.

In conclusion, our anisotropy analysis of LaCrGe$_3$ does not reveal any anomalies that would cause the re-opening of hysteresis loops or the shark-fin shape of the $M$\,v\,$T$ curves.

%\bibliographystyle{apsrev4-2.bst}
%\bibliography{references,biblio}% Produces the bibliography via BibTeX.

%

\end{document}